\documentclass[12pt]{article}
\usepackage{amssymb}
\usepackage{amsmath}
\usepackage[font={small,it}]{caption}
\usepackage{tikz}
\usetikzlibrary{shapes.geometric}
\usetikzlibrary{arrows,matrix,calc,scopes,decorations.markings}
\usepackage[nodisplayskipstretch]{setspace}
\usepackage{setspace}
\usepackage{amsfonts}
\usepackage{fancyhdr}
\usepackage{xcolor}
\usepackage{caption}
\usepackage{graphicx}
\usepackage[percent]{overpic}
\usepackage{rotating}
\usepackage{comment}
\usepackage{color}
\usepackage{cite}
\usepackage{braket}
\definecolor{darkgreen}{rgb}{0,0.5,0}
\definecolor{darkblue}{rgb}{0,0,0.6}
\definecolor{purple}{rgb}{0.4,.2,0.7}

\newcommand{\be}{\begin{equation}}
\newcommand{\ee}{\end{equation}}

\usepackage[colorlinks=true,citecolor=darkgreen,linkcolor=black,urlcolor=purple]{hyperref}

\usepackage{pdfsync}

\makeatletter
\newcommand*{\defeq}{\mathrel{\rlap{%
                     \raisebox{0.3ex}{$\m@th\cdot$}}%
                     \raisebox{-0.3ex}{$\m@th\cdot$}}%
                     =} 
\makeatother

\def\be{\begin{eqnarray}}
\def\ee{\end{eqnarray}}

\newcommand{\bea}{\begin{eqnarray}}
\newcommand{\eea}{\end{eqnarray}}
\def\ben{\begin{equation}}
\def\een{\end{equation}}

     \let\r=v

\def\be{\begin{equation}}
\def\ee{\end{equation}}
\def\ba{\begin{array}}
\def\ea{\end{array}}

\def\ba#1\ea{\begin{align}#1\end{align}}
\def\bs#1\es{\begin{split}#1\end{split}}

\newcommand\blfootnote[1]{%
  \begingroup
  \renewcommand\thefootnote{}\footnote{#1}%
  \addtocounter{footnote}{-1}%
  \endgroup
}

\allowdisplaybreaks  

\begin{document}
\doublespacing

\begin{center}

~
\vskip5mm

{\Large  \textbf{{
Building up quantum spacetimes with BCFT Legos}}}

\vskip7mm

Ling-Yan Hung${}^{1,2,\dagger}$ \blfootnote{$\dagger$  lyhung@tsinghua.edu.cn}and Yikun Jiang${}^{3,*}$\blfootnote{$\dagger$  phys.yk.jiang@gmail.com}

\vskip5mm

\it{${}^1$ Yau Mathematical Sciences Center, Tsinghua University, Beijing 100084, China}

\it{${}^2$ Yanqi Lake Beijing Institute of Mathematical Sciences and Applications (BIMSA), Huairou District, Beijing 101408, China}

\it{${}^3$ Department of Physics, Northeastern University, Boston, MA 02115, USA}

\vskip5mm

\end{center}

 \vspace{4mm}

\begin{abstract}
\noindent

Is it possible to read off the quantum gravity dual of a CFT directly from its operator algebra? In this essay, we present a step-by-step recipe synthesizing results and techniques from conformal bootstrap, topological symmetries, tensor networks, a novel symmetry-preserving real-space renormalization algorithm devised originally in lattice models, and the asymptotics of quantum $6j$ symbols, thereby providing  an answer in the affirmative. Quantum 2D Liouville theory serves as a simple and explicit example, illustrating how the quantum gravitational path integral can be built up from local pieces of BCFT correlation functions, which we call the ``BCFT Legos''. The constructive map between gravity and CFT naturally and explicitly bridges local geometrical data, algebraic structures, and quantum entanglement, as envisaged by the {\it It from Qubit} motto.  

 \vspace{1.5 cm}
{\centering
\emph{Expanded version of essay written for the Gravity Research Foundation 2024 Awards for Essays on Gravitation.}}
 \end{abstract}

\pagebreak
\pagestyle{plain}
\setcounter{tocdepth}{2}{}
\vfill

\clearpage
\pagenumbering{arabic} 
\section{Introduction}
The holographic gauge-gravity correspondence in its strongest form conjectures the equivalence between a theory of quantum gravity and a quantum field theory (QFT) in the absence of gravity in one lower dimension\cite{tHooft:1993dmi, Susskind:1994vu, Maldacena:1997re}. Should this correspondence be true, it implies that the {\it geometries} are encoded in the {\it algebraic data} of the QFT. There are numerous hints of such an encoding -- most notably the ``ER=EPR'' proposal in the context of the AdS/CFT correspondence\cite{Maldacena:1997re, Witten:1998qj, Gubser:1998bc}, which is the manifestation of CFT entanglement as geometrical connectivity in the bulk, a profound observation encapsulated in the {\it It from Qubit} motto\cite{Wheeler1989-WHEIPQ, Maldacena:2001kr, Ryu:2006bv, VanRaamsdonk:2009ar, VanRaamsdonk:2010pw, Maldacena:2013xja, VanRaamsdonk:2018zws}. 

One prominent difficulty to make progress is that, in most cases, we lack a non-perturbative understanding or even a definition for either side of the correspondence, which is more formidable than purely an issue of tractability of computations. For example, as far as quantum gravity is concerned, we generally do not have a precise definition of the measure for the path integral, let alone the ability to compute it.  It is thus difficult to formulate this equivalence beyond perturbative limits in the central charge $c$ or Newton's constant $G_N$.

To make progress, we looked elsewhere -- the development of another type of holographic correspondence, bridging QFT with topological symmetries (see \cite{Schafer-Nameki:2023jdn, Shao:2023gho} and references therein) and a TQFT in one higher dimensions \cite{cat1,cat2,cat3}. Several families of examples showing such correspondence have been known for a long time, notably a 2D rational CFT (RCFTs) is related to some 3D TQFT characterised by the Moore-Seiberg tensor category $\mathcal{C}$ associated to that CFTs\cite{Moore:1988qv}. Since the seminal work\cite{Witten:1988hf}, it has been well known  that the space of conformal blocks on a 2D surface is isomorphic to the Hilbert space of a 3D TQFT. In the modern language of generalized symmetries, the TQFT is capturing the topological symmetries generated by Verlinde lines of the RCFT\cite{Verlinde:1988sn, Chang:2018iay}. Moreover, the full CFT data, including its modular invariant spectrum, is encoded in a module category $\mathcal{M}_\mathcal{C}$ of $\mathcal{C}$\cite{Fuchs:2002cm, Fuchs:2004xi}. It is proposed that the {\it full} RCFT path integral can be constructed from a 3D ``sandwich'' - one performs the path integral of the TQFT on a 3D slab with topological boundary condition characterised by $\mathcal{M}_\mathcal{C}$ on one boundary, and some non-trivial (but yet to be understood) conformal boundary condition on the other\cite{Gaiotto:2020iye}. The $\mathcal{M}_\mathcal{C}$ topological boundary basically instructs the TQFT on what linear combination of conformal blocks should be included in the path integral. To realize the sandwich explicitly, we leveraged an observation from the condensed matter community, that recovers 2D lattice integrable models from giving non-trivial boundary conditions to the Turaev-Viro state-sum formulation of the 3D TQFT that triangulates the 3D bulk\cite{Turaev:1992hq, Frank1, Aasen:2020jwb}. It turns out that the TQFT also teaches one how to perform a 2D real space renormalization group (RG) flow analogous to ``spin blocking"\cite{PhysicsPhysiqueFizika.2.263}, in a specific way that protects the topological symmetries. When one follows through the RG flow to a fixed point, it recreates the exact CFT path integral, realizing the sandwich as an explicit state-sum. This whole procedure can be done analytically in an exact manner, relying on the triple roles (to be explained below) of the local building blocks which we call boundary conformal field theory (BCFT) Legos\cite{Cardy:2004hm, Takayanagi:2011zk, lego, Cao:2021ibt, 1998math......9057B} in this paper. 
 This idea is developed for RCFTs in \cite{Chen:2022wvy, Cheng:2023kxh}.

This picture provides a fresh perspective on the holographic correspondence between CFT$_2$ and AdS$_3$ gravity which is also topological\cite{Achucarro:1986uwr, Witten:1988hc, Verlinde:1989ua, Carlip:1991zm, Coussaert:1995zp, Collier:2023fwi}. As an example, we apply the above framework to Liouville CFT\cite{Chen:2024unp}. Overcoming technical hurdles coming with an irrational CFT, we get an explicit, computable and non-perturbative quantum measure for a sum over 3D hyperbolic geometries. The result can be written in the form of an explicit ``tensor network'' that was proposed to be relevant for the emergence of  geometries\cite{Swingle:2009bg, VanRaamsdonk:2018zws}. Not only does it provide a rare example of a precise quantum measure for geometries, it illustrates the encoding of {\it geometries} in {\it algebraic data} that lies at the heart of the AdS/CFT correspondence. Moreover, this encoding is {\it local} - the quantum AdS geometry is built up piecewise from local BCFT correlation functions, epitomising the {\it It from Qubit} principle. 

In the following, let's go through the recipe of building up quantum spacetimes with BCFT Legos!

\section{Step 1: Exact CFT path integral from triangulation}
    \begin{figure}[h]
\vspace{-0.5 cm}
\begin{minipage}[b]{0.47\linewidth}
\begin{overpic}[scale=0.18]{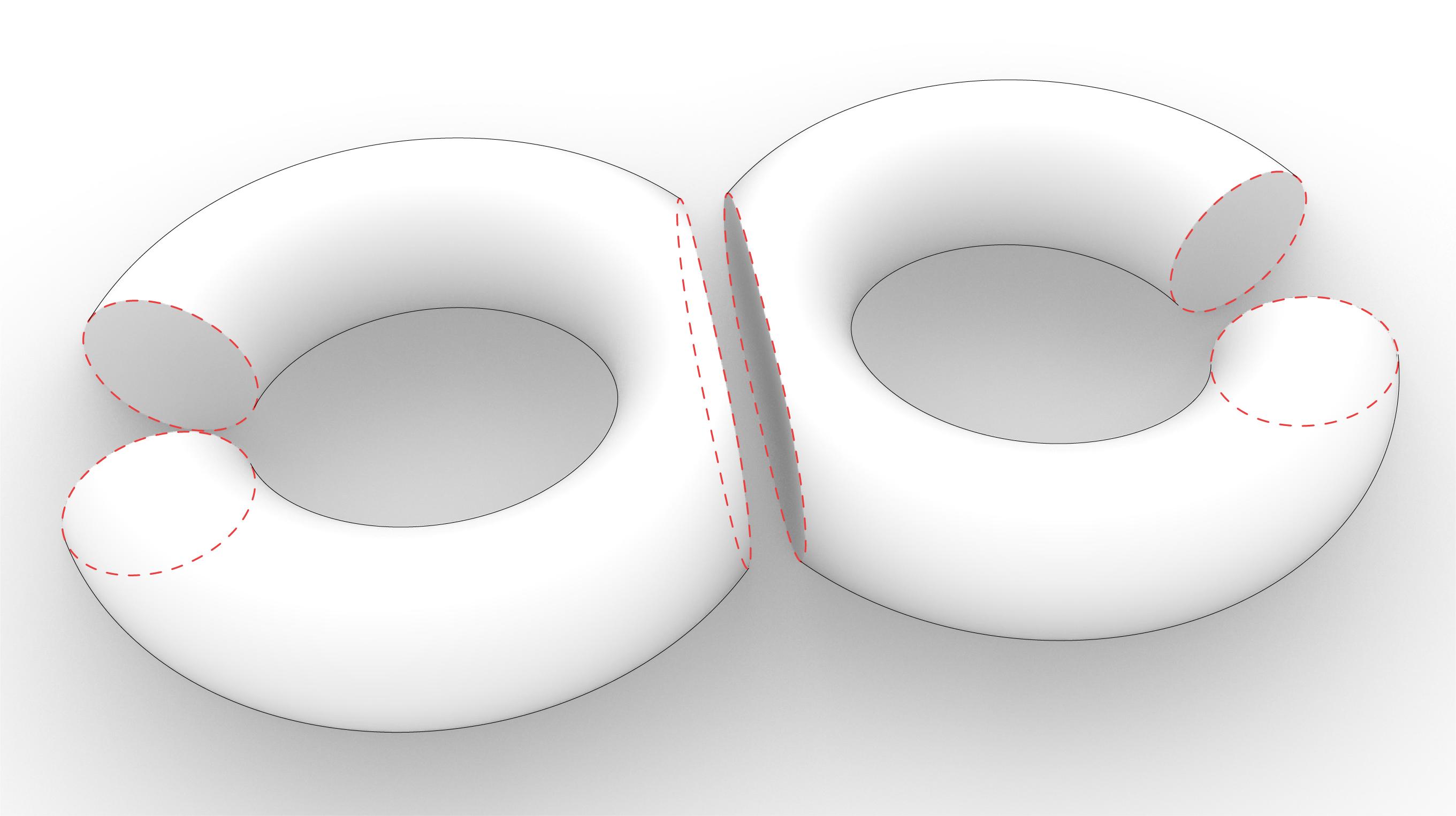}
\end{overpic} 
        \end{minipage}
        \begin{minipage}[b]{0.47\linewidth}
\begin{overpic}[scale=0.18]{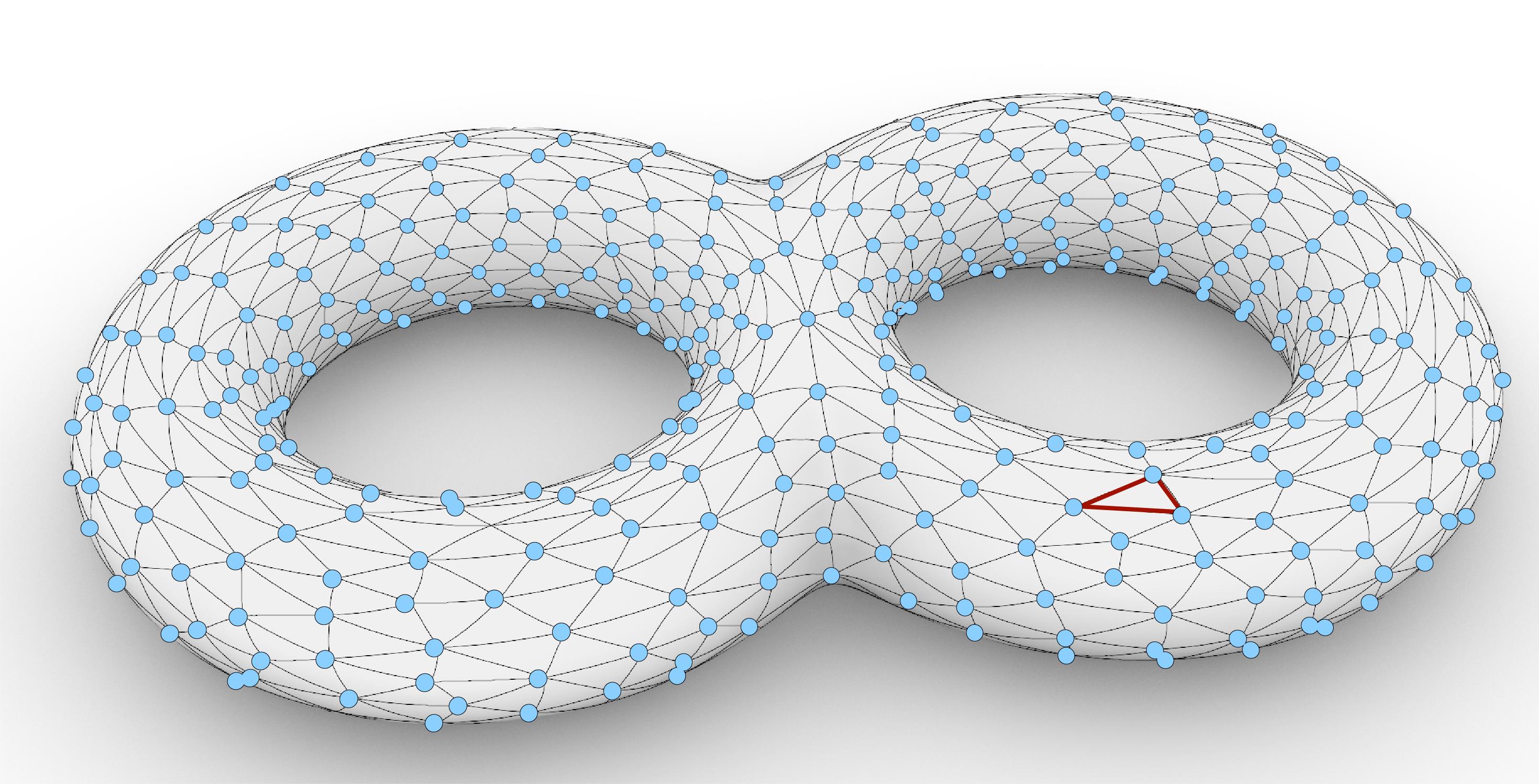}
\end{overpic}  
  \end{minipage}
\caption{Two different cutting-and-gluing procedures for the calculation of CFT genus two partition function. (Left): Two pairs of pants are glued along three circles. (Right): Arbitrarily many chipped triangles are glued together along common edges, with boundary conditions specified on the tiny blue holes. The exact CFT partition function is reproduced when the holes are contractible.}\label{cutting and gluing}
\end{figure}

\begin{figure}
	\centering
	\includegraphics[width=0.45\linewidth]{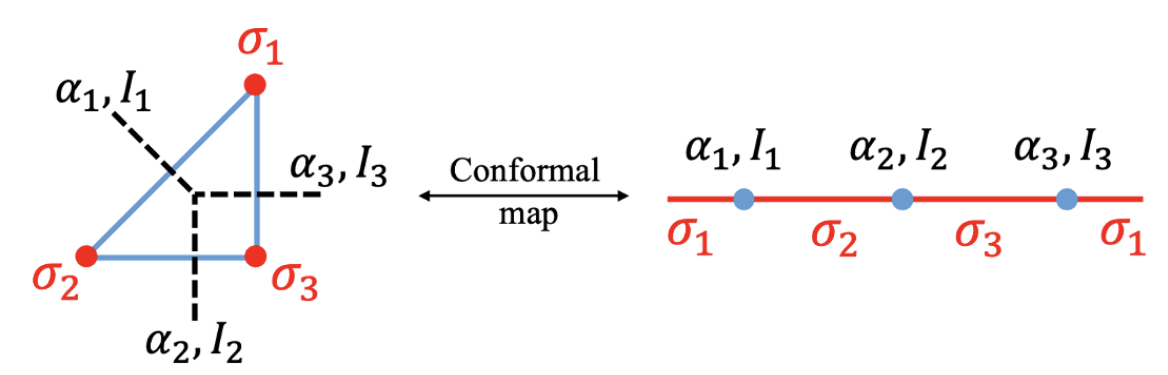}
	\caption{BCFT path integral on a triangle is calculated by conformally mapping it to the upper half plane.}
	\label{conformalthreepoint}
\end{figure}

Equipped with Virasoro symmetry, the conventional method for calculating the exact 2D CFT path integral on a genus $g$ closed Riemann surface with $n$ operator insertions involves cutting the 2D manifold into $2g-2+n$ pairs of pants, then sewing them back together by summing over the complete basis of states defined on shared circles. In this paper, we will consider another new type of cutting-and-gluing procedure \cite{Cheng:2023kxh, Chen:2024unp}. This method disassembles the 2D manifold into triangles, which serve as the basic building blocks capable of re-enacting path integral on any surface and we refer to the quantities associated to them as the ``BCFT Legos''. Figure \ref{cutting and gluing} compares the two different cutting-and-gluing procedures related to the calculation of the genus two partition function. For regularization purposes, we slightly chip the corners of the triangle. At each chipped corner, we specify a conformal boundary condition\cite{Cardy:2004hm} labeled $\sigma_i$, and along each edge (instead of a circle), we define a complete basis of states, with $\alpha_i$ labeling the primary conformal dimensions, and $I_i$ the descendants. Each triangle can be conformally mapped to the upper half plane as shown in Figure \ref{conformalthreepoint}. By conformal invariance, we can express the contribution of each triangle as:
\be
\mathcal{T}^{\sigma_1\sigma_2\sigma_3}_{(\alpha_1, I_1) (\alpha_2, I_2)(\alpha_3, I_3)}(\triangle) = C^{\sigma_3\sigma_1\sigma_2}_{\alpha_1 \alpha_2\alpha_3} \gamma^{\alpha_1 \alpha_2\alpha_3}_{I_1I_2I_3}(x_i, \epsilon)
\ee
where $C^{\sigma_3\sigma_1\sigma_2}_{\alpha_1 \alpha_2\alpha_3}$ is the BCFT three-point structure coefficient, and $\gamma^{\alpha_1 \alpha_2\alpha_3}_{I_1I_2I_3}(x_i, \epsilon)$ is the function that encodes the positions of the end points $x_i$, the size of the holes $\epsilon$, and the shape of the triangle. The sum over descendants $I_i$ on shared edges for adjacent triangles provides the usual definition of conformal blocks, and we will just call $\gamma^{\alpha_1 \alpha_2\alpha_3}_{I_1I_2I_3}(x_i, \epsilon)$ the conformal blocks for simplicity. The path integral over the whole 2D manifold $Z(\sigma_i)$ involves a product over all triangles and a sum over intermediate states $\alpha_i$ and $I_i$.
Notice that $Z(\sigma_i)$ depends on our choice of boundary conditions $\sigma_i$ fixed at the corners. 

To reproduce the exact CFT path integral, we need to eliminate the dependence on the boundary conditions and close up all the tiny holes without leaving any imprint. To achieve this, we follow the prescription proposed in \cite{Hung:2019bnq, Brehm:2021wev, Cheng:2023kxh, Chen:2024unp} where a weighted sum over boundary conditions is performed. The weights $\omega_i$ are chosen to be proportional to the quantum dimensions $d_i$ associated to the boundary conditions $\sigma_i$. This weighted sum of $\omega_i$ projects onto the vacuum modules in the closed CFT channel \cite{Cheng:2023kxh, Chen:2024unp}. In the limit of a shrinking size $\epsilon \to 0$, only the vacuum state gives a universal divergent contribution $e^{\frac{\pi c}{6 \epsilon}} $, from the Casimir energy and all descendant contributions are suppressed. This means that the holes become contractible, and we can reproduce the exact CFT path integral by shrinking all the holes on the vertices $v$,
\be \label{Zproduct}
Z=\lim_{\epsilon \to 0} \sum_{\sigma_i} \prod_v \left(\omega_{v(\sigma_i)} e^{-\frac{\pi c}{6 \epsilon}}\right) \sum_{\alpha_i}  \sum_{\{I_i\}}     \prod_{\triangle} \mathcal{T}^{\sigma_1\sigma_2\sigma_3}_{(\alpha_1, I_1) (\alpha_2, I_2)(\alpha_3, I_3)}(\triangle)
\ee

Hense, we successfully build up the exact 2D CFT path integral using BCFT Legos.

\section{Step 2: CFT from topological symmetry and 3d TQFT}

In fact, the method outlined above for constructing CFT partition functions automatically lends itself a 3D TQFT interpretation. To make the connection most explicit, we first rescale $C^{\sigma_3\sigma_1\sigma_2}_{\alpha_1 \alpha_2\alpha_3}$ and $\gamma^{\alpha_1 \alpha_2\alpha_3}_{I_1I_2I_3}$ simultaneously, moving to the ``Racah gauge" \cite{Kojita:2016jwe, Kojita:2016jwe, Cheng:2023kxh, Chen:2024unp}. 

\begin{figure}[htp]
\centering
\includegraphics[width=.32\textwidth]{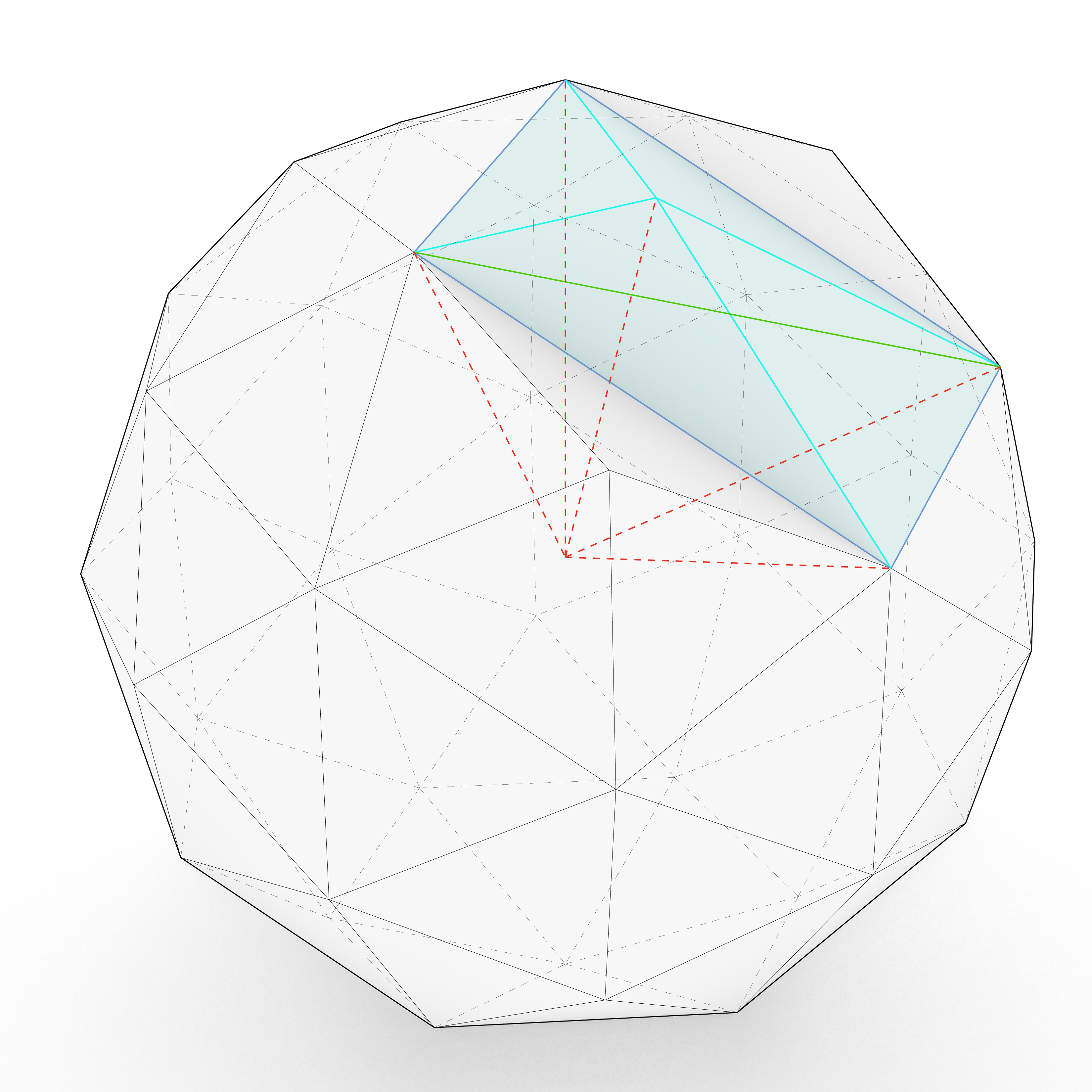}\hfill
\includegraphics[width=.35\textwidth]{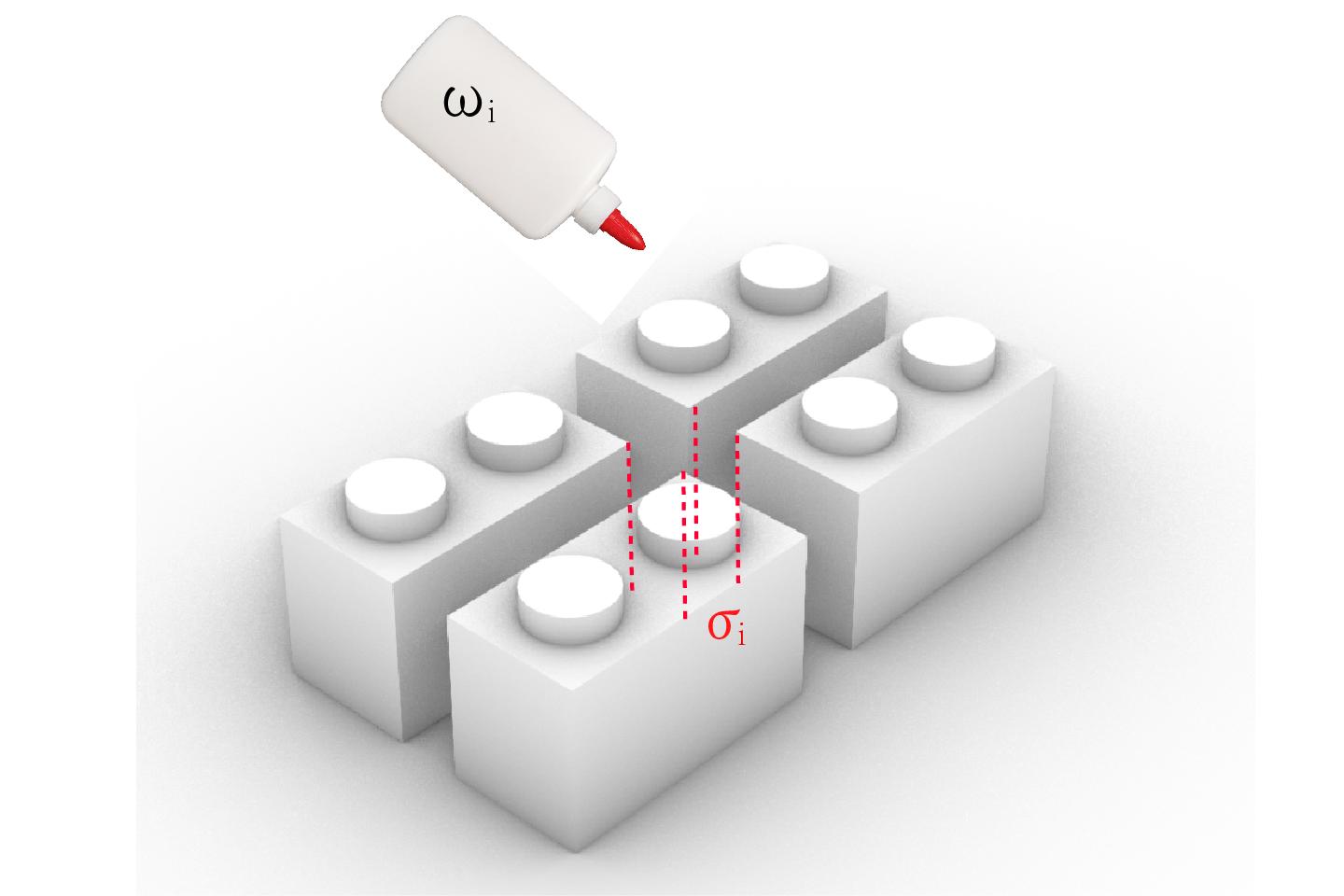}\hfill
\includegraphics[width=.32\textwidth]{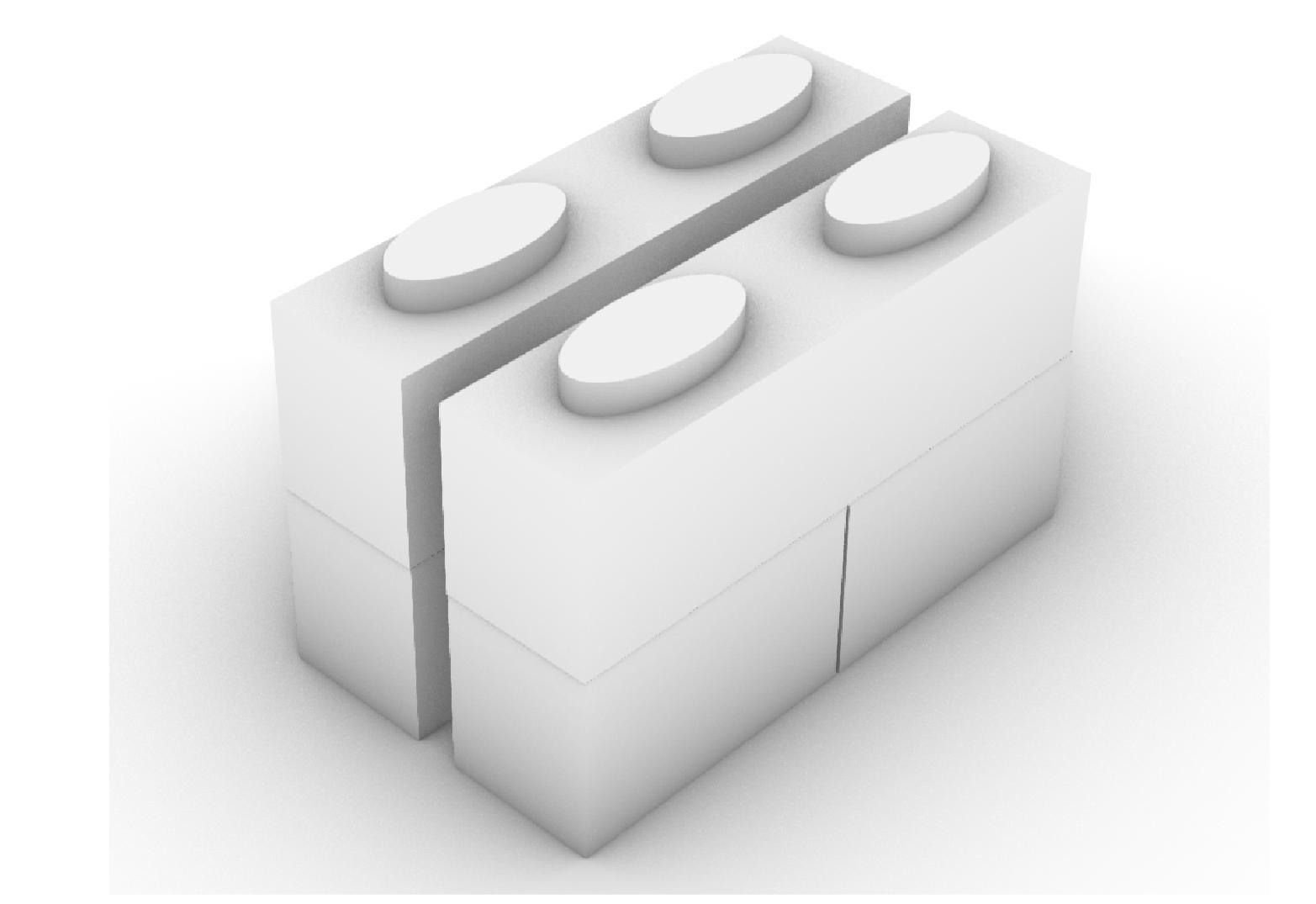}
\caption{(Left): The 3D geometrical picture corresponding to the 2D coarse-graining in Figure \ref{reduce_bubble}. The state, locally defined on the four cyan edges, contracts the indices with the $6j$ symbol tensors corresponding to the shaded tetrahedra, and is mapped to a state defined on a single green edge. (Right): Larger Lego blocks from “spin-blocking” smaller ones.}
\label{azure}
\end{figure}

In this gauge, the structure coefficients manifest their \textbf{second role}, as they are proportional to the quantum $6j$ symbols for the Moore-Seiberg tensor category $\mathcal{C}$ associated with the CFT\cite{Moore:1988qv}\footnote{$\hat{C}^{\sigma_3\sigma_1\sigma_2; \text{Racah}}_{\alpha_1 \alpha_2\alpha_3}=\left(d_{\alpha_1} d_{\alpha_2} d_{\alpha_3}\right)^{1/4} \begin{Bmatrix}
\alpha_1 & \alpha_2 & \alpha_3\\
\sigma_3 & \sigma_1 &  \sigma_2
\end{Bmatrix}$}. Then, we can express the CFT partition function as an overlap, or a ``strange correlator''\cite{Frank1, Chen:2022wvy, Cheng:2023kxh, Chen:2024unp}\footnote{$\langle\Omega|=  \sum_{\alpha_i, I_i} \langle\{\alpha_i \}|\prod_{\triangle} \hat \gamma^{\alpha_1 \alpha_2 \alpha_3}_{I_1I_2I_3}, \qquad |\Psi \rangle= \prod_v  \sum_{\sigma_i,\alpha_i} \sqrt{d_{\alpha_i}} \left( \omega_{v(\sigma_i)} e^{-\frac{\pi c}{6 \epsilon}}\right)\prod_{\triangle} 
\begin{Bmatrix}
    \alpha_1 & \alpha_2 & \alpha_3  \\
    \sigma_3 & \sigma_1 & \sigma_2
    \end{Bmatrix} |\{\alpha_i\}\rangle$}: 
\vspace{-0.2 cm}
\be \label{overlap}
Z=\lim_{\epsilon\to 0} \langle \Omega | \Psi \rangle 
\vspace{-0.2 cm}
\ee
In this expression, both states are defined on the edges of the triangles. The $\bra{\Omega}$ state is comprised of a product of the conformal blocks $\gamma^{\alpha_1 \alpha_2\alpha_3}_{I_1I_2I_3}(x_i, \epsilon)$, as indicated in \eqref{Zproduct}, while the state $\ket{\Psi}$ originates from the remainder. Upon expressing these in terms of the $6j$ symbols, it becomes apparent that the state $\ket{\Psi}$ embodies a ground state of the Levin-Wen string net model\cite{Levin:2004mi}, or equivalently, the 3d Turaev-Viro state-sum TQFT\cite{Turaev:1992hq} based on $\mathcal{C}$! Each triangle on the 2D boundary is associated with a tetrahedron in the 3D bulk, each being assigned a $6j$ symbol for the 3D state-sum. The chosen weights $\omega_i$ are naturally present in the Turaev-Viro TQFT, and are attributed to each 3D internal edge within the state-sum. The internal edges, as shown in Figure \ref{azure} are depicted in red.
\begin{figure}
	\centering
	\includegraphics[width=0.34\linewidth]{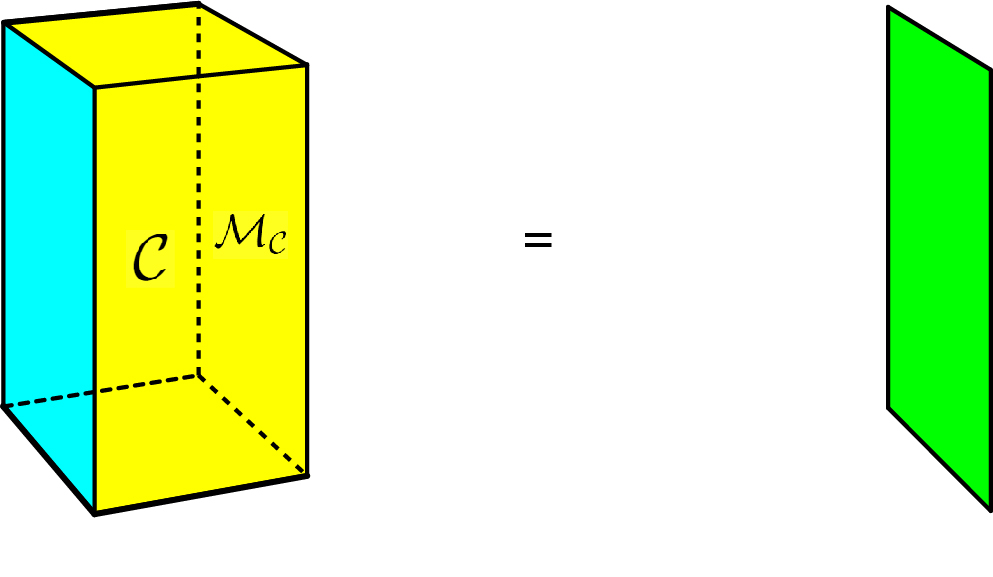}
	\caption{The sandwich construction for the full 2D CFT partition function from 3D TQFT. In 3D TQFT, the left (blue) boundary is the conformal boundary, and the right (yellow) one is the topological boundary. Collapsing the 3D bulk, the 2D theory lives on the green(=blue+yellow) surface.} \label{sandwich}
\end{figure}
 
The overlap \eqref{overlap} precisely embodies the ``sandwich construction'' depicted in Figure \ref{sandwich}\cite{Gaiotto:2020iye}. It was suggested that the path integral of a QFT with topological symmetries can be understood as the path integral of a TQFT in one higher dimension on a thin slab, flanked by a topological and a non-topological boundaries. The topological symmetries made explicit by the 3D theory are the 2D Verlinde lines which feature as Wilson lines in 3D\cite{Verlinde:1988sn}. The ground state $\ket{\Psi}$, formed from the $6j$ symbols, arises from a TQFT path integral with a topological boundary condition on one side \cite{Lootens:2020mso}. The data regarding the 3D topological boundary is hidden in the set of 2D conformal boundary conditions\cite{Cardy:2004hm} of the CFT, labeled by objects in a module category $\mathcal{M}_{\mathcal{C}}$ of ${\mathcal{C}}$,\footnote{For diagonal CFTs, $\mathcal{M}_{\mathcal{C}} = {\mathcal{C}}$ \cite{Fuchs:2002cm, Fuchs:2004xi}, obscuring the presence of the topological boundary.}and encompasses all data pertinent to the full modular invariant CFT \cite{Fuchs:2002cm, Fuchs:2004xi}. The complete correspondence between 3D TQFT and  full 2D CFT involves both the isomorphism between the Hilbert space and the space of conformal blocks, and {\it the module category $\mathcal{M}_{\mathcal{C}}$, which selects the modular invariant spectrum for the CFT}. Meanwhile, our $\bra{\Omega}$, composed of conformal blocks establishes a conformal boundary condition for the slab. The overlap thus perfectly synthesizes the 2D and 3D theories.

\section{Step 3: 3D tensor network from 2D Real space RG flow}

We have demonstrated that the overlap \eqref{overlap} encodes a 3D bulk in a sandwich, but the bulk takes up a triangulation with a single layer of tetrahedra.  Utilizing the freedom in changing the triangulation within a TQFT, we now demonstrate the capability of growing more and more layers to probe deeper and deeper into the bulk by switching triangulation systematically, thus producing an exact ``tensor network'' that implements an RG procedure in the CFT\cite{Swingle:2009bg}.

The idea is to use the gluing rules for the $6j$ symbols, more specifically the pentagon identity
which connects two tetrahedra to three, and the orthogonality condition that eliminates a pair of tetrahedra\cite{Chen:2022wvy, Cheng:2023kxh, Chen:2024unp}. Figure \ref{azure} illustrates an example, where a state defined on a lattice can be locally represented as a tensor with four uncontracted indices (the four cyan edges), and is mapped to another state with a single-index tensor (the ``coarse grained'' green edge), through contractions provided by the $6j$ symbols (the two shaded tetrahedra). This local move can be repeated across the entire lattice to change the lattice spacing from $\Lambda$ to $\Lambda' = \sqrt{2} \Lambda$, resulting in a linear map $U$ such that $|\Psi\rangle_{\Lambda} = U |\Psi\rangle_{\sqrt{2} \Lambda}$ ($U$ is the
collection of $6j$ symbols associated to the shaded tetrahedra). The overlap ${}_{\Lambda}\langle \Omega|\Psi\rangle_{\Lambda} = {}_{\Lambda}\langle \Omega| U |\Psi\rangle_{\sqrt{2} \Lambda} $ suggests that we can absorb $U$ into ${}_{\Lambda}\langle \Omega|$ to obtain the coarse-grained state
\be
_{\sqrt{2}\Lambda}\bra{\Omega}= _{\Lambda}\bra{\Omega} U
\ee
The linear map $U$ is thus changing the triangulation scale and performing a step of real space RG flow on the state $\bra{\Omega}$  \cite{Buerschaper_2009,Gu_2009,Frank1, Chen:2022wvy, Cheng:2023kxh, Chen:2024unp}, essentially an example of ``spin-blocking''\cite{PhysicsPhysiqueFizika.2.263}! 

We can start from any state ${}_{\Lambda}\bra{\Omega}$ and run this RG procedure. Typically, the fixed point is a state corresponding to a 2D TQFT. However, through fine tuning the initial state, we can land on a phase transition fixed point between different 2D TQFTs, and that's where 2D CFTs reside\cite{Chen:2022wvy}. 

\begin{figure}
	\centering
	\includegraphics[width=0.8\linewidth]{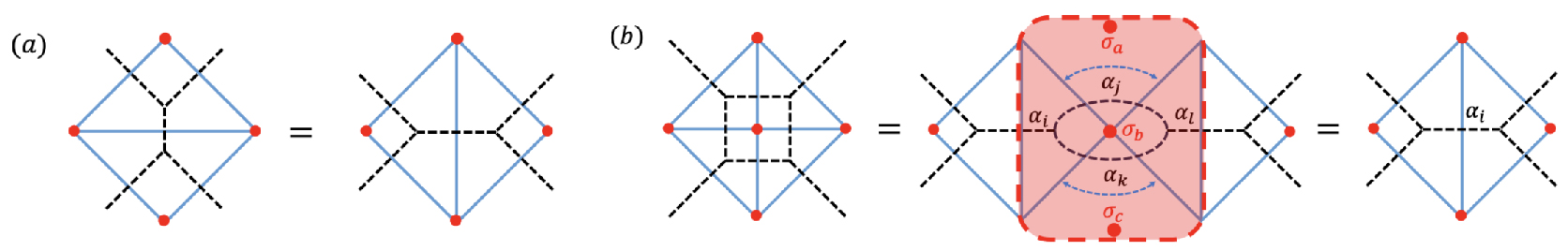}
	\caption{(a). Graphical representation of the local crossing relation. (b). Example of a real space RG step on the 2D boundary. We use crossing relation twice for the first equation, and the second equation comes from the pentagon identity and orthogonality condition. The lattice scale on the right is $\sqrt{2}$ times larger than the one before the RG move.}
	\label{reduce_bubble}
\end{figure}

The $\langle \Omega |$ in (\ref{overlap}) comes from products of conformal blocks, thus explicitly takes the form following from a CFT path integral, and is a fixed point of the flow operator $U$. The most crucial ingredient in checking this assertion directly is to use the \textbf{third role} played by the $6j$ symbols: they are also related to the conformal crossing kernels relating conformal blocks in different channels,
\be
\vcenter{\hbox{
\begin{tikzpicture}[scale=0.5]
\draw (0,1) -- (1, 0);
\draw (0, -1) -- (1,0);
\draw (1,0) -- (3,0);
\draw (3,0) -- (4,1);
\draw (3,0) -- (4,-1);
\node at (-0.5,1.25) {$\alpha_1$};
\node at (-0.5,-1.25) {$\alpha_2$};
\node at (4.5,1.25) {$\alpha_3$};
\node at (4.5,-1.25) {$\alpha_4$};
\node at (2,0.6) {$\beta_1$};
\end{tikzpicture}
}}
= \sum_{\beta_2} F_{\beta_1,\beta_2} \begin{pmatrix}
\alpha_1 & \alpha_3 
\\
\sigma_2 & \sigma_4 
\end{pmatrix}
\vcenter{\hbox{
\begin{tikzpicture}[scale=0.5]
\draw (0,0) -- (1,-1);
\draw (1,-1) -- (2,0);
\draw (1,-1) -- (1,-3);
\draw (1,-3) -- (0, -4);
\draw (1,-3) -- (2,-4);
\node at (-0.5,0.25) {$\alpha_1$};
\node at (2.5,0.25) {$\alpha_3$};
\node at (-0.5, -4.25) {$\alpha_2$};
\node at (2.5, -4.25) {$\alpha_4$};
\node at (1.6,-2) {$\beta_2$};
\end{tikzpicture}
}}
\ee
In the Racah gauge, they are again proportional to the $6j$ symbols\footnote{$F^{\text{Racah}}_{\sigma_2,\beta_3} \begin{pmatrix}
\beta_2 & \beta_1 
\\
\sigma_3 & \sigma_1 
\end{pmatrix}=\sqrt{d_{\beta_3} d_{\sigma_2}}
\begin{Bmatrix}
\beta_2 & \beta_1 & \beta_3\\
\sigma_1 & \sigma_3 & \sigma_2
\end{Bmatrix}$}. 
The crossing invariance of the overlap is verified locally by merging the pentagon identity for the $6j$ symbols with the properties of the crossing kernels, as demonstrated in Figure \ref{reduce_bubble} (a).

We can now consider diagrams like Figure \ref{reduce_bubble} (b). The middle diagram is obtained from the left by applying local crossing invariance twice. Through direct calculation of the middle diagram and leveraging properties of crossing kernels, we obtain the right diagram, which explicitly affirms the contractibility of the holes mentioned in Step 1. The procedure in going from the left diagram to the right diagram is precisely the one that changes the triangulation scale, in which $|\Psi\rangle_{\Lambda}$ is replaced by $U |\Psi\rangle_{\sqrt{2}\Lambda}$, and subsequently $U$ is absorbed by ${}_\Lambda\langle \Omega|$ to give  ${}_{\sqrt{2}\Lambda}\langle \Omega|$, which is locally identical to ${}_\Lambda\langle \Omega|$.   In other words, ${}_\Lambda\langle \Omega|$ is a fixed point of the real space RG generated by $U$ as expected.

\begin{figure}
	\centering
	\includegraphics[width=0.5\linewidth]{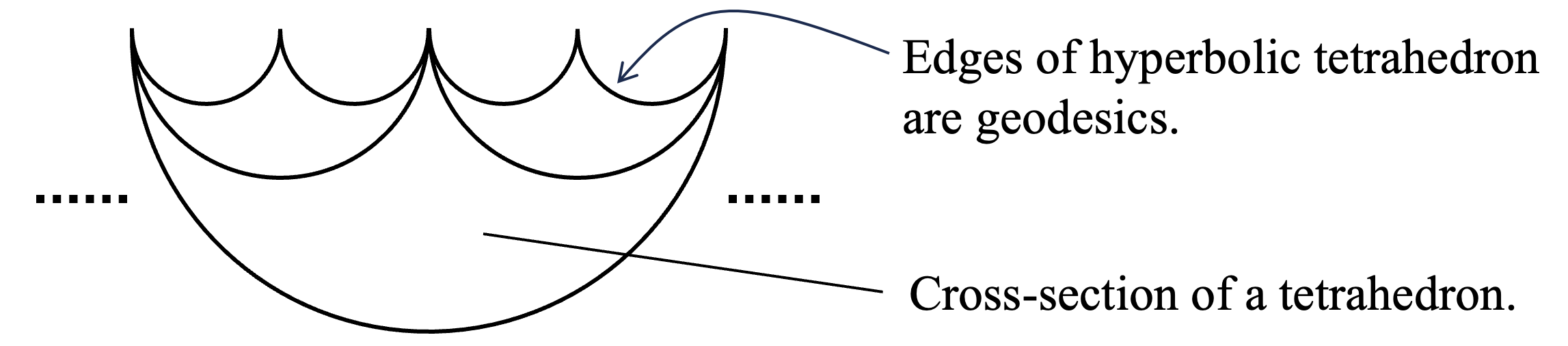}
	\caption{The 2D cross-section of several consecutive RG moves $U^N(\lambda)$.}
	\label{crosssection}
\end{figure}

We can keep running the coarse-graining procedure, using $|\Psi\rangle_{\Lambda} = U^N |\Psi\rangle_{\sqrt{2}^N\Lambda}$, for the state $\bra{\Omega}$ defined on the 2D boundary. The 3D bulk is thus converted into layers of tetrahedra geometrically forming the RG operator $U^N$, and each layer down probes deeper into the bulk. Figure \ref{crosssection} displays a cross-section of $U^N$. This is a precise realization of a 3D bulk tensor network stemming from the 2D CFT boundary\cite{Swingle:2009bg}, 
and it geometrically realizes the UV/IR correspondence in AdS/CFT in an exact fashion! Notice that for each RG step, we are adding entanglement to the state $\bra{\Omega}$ through increasing the bond dimension\cite{Verstraete:2004cf, 2017PhRvB..96c5101L, Chen:2022wvy}. Ultimately, we reach the fixed point $\bra{\Omega}$ state, which stays invariant under further action of $U$ since the bond dimensions are already infinity(populated by the infinite number of descendants). 

In summary, we have employed our Legos to construct $U^N$, a ``pregeometry''  that looks like what we anticipate for the AdS bulk in AdS/CFT correspondence. The final step is to show that this bulk indeed represents some AdS geometry. In fact, since the CFT is a quantum theory, we want the bulk to be a sum over quantum geometries so that it can be regarded as defining some {\it Quantum Gravity}.
\section{Step 4: 3D tensor network is Quantum Gravity}
We will now demonstrate that the 3D tensor network we have constructed is indeed Quantum Gravity. We start with the explicit simple example of Liouville theory\cite{Li:2019mwb, Chen:2024unp}, which offers an unconventional 3D gravitational theory comprising  solely gravitons, whose exact meaning will be specified below.

As a theory with a continuous physical spectrum, there are extra subtleties related to convergence that we address comprehensively in \cite{Chen:2024unp}. With meticulous attention to these issues and a carefully selected normalization for the Liouville operators, the same construction above is proven effective for Liouville CFT. The sums we illustrated above are all replaced by integrals, and Liouville theory has a flat spectrum in Liouville momentum $P_{\alpha} \in \mathbb{R}^+$, so we have, $\sum_{\alpha_i} \to \int_0^\infty dP_{\alpha_i}$.
Quantum dimensions and $6j$ symbols are substituted with the Plancherel measure\footnote{The importance of the Plancherel measure in the study of 3D gravity has been emphasized in \cite{McGough:2013gka, Jackson:2014nla, Mertens:2022ujr, Chua:2023ios}.} and the $b$-$6j$ symbol for a special representation, known as the ``modular double'' of the quantum group $\mathcal{U}_q(SL(2,\mathbb{R}))$, where $q$ is linked to $c$\cite{Faddeev:1999fe, Ponsot:1999uf, Teschner:2012em}. The formulas are for any finite values of $c$, hence a matching with the semi-classical gravity behavior in the large $c$ (or small $G_N$) limit would provide a non-perturbative UV completion for the resultant gravitational theory.

Simplification arises in 3D as any solution to Einstein equation without matter is locally diffeomorphic to AdS$_3$. Therefore, our focus is to ensure that the b-$6j$ symbols assigned to the tetrahedra in our bulk construction match with some Einstein-Hilbert action on AdS$_3$. The physical meaning of the boundary primary labels $\alpha_i$ (or equivalently $P_{\alpha_i}$) will also be understood from the matching.

At leading order in $G_N$, the Einstein-Hilbert action is directly proportional to the 3D volume,
\be
S_{EH} =- \frac{1}{16\pi G_N}\int_{H} d^3 x \sqrt{g} (R - 2\Lambda) =  \frac{V(H)}{4\pi G_N},
\ee
and remarkably, the large $c$ limit of the b-$6j$ symbols gives\cite{Murakami1, Teschner:2012em, Chen:2024unp},
 \be \label{eq:classlim2}
\begin{Bmatrix}
\alpha_4& \alpha_5 &\alpha_6 \\
\alpha_1 & \alpha_2 & \alpha_3
    \end{Bmatrix}_b    \xrightarrow[]{b\to 0}\exp{ \left(- \frac{ \textrm{Vol}(T\{l_i\}) + \sum_i l_i \theta_i/2   }{\pi b^2}  \right)}  
\ee

where $b$ is related to the central charge via $c \approx 6/b^2$, $\textrm{Vol}(T\{l_i\})$ is the volume of the 3D hyperbolic tetrahedron whose geodesic edge lengths are $l_i$, and $\theta_i$ are the dihedral angles between the faces. The identification $l_i=2\pi b P_{\alpha_i}$ manifests the physical meaning of the Liouville momenta $P_i$ as geodesic lengths. The last term in \eqref{eq:classlim2} is the Hayward corner term introduced along the codimension-two edges to ensure a well-defined variational problem in gravity\cite{PhysRevD.47.3275, Takayanagi:2019tvn}. So the large $c$ limit of the b-$6j$ symbols matches with the gravitational on-shell action once we identify $c =\frac{3}{2 G_N}$, echoing the celebrated relation in the AdS$_3$/CFT$_2$ correspondence \cite{Brown:1986nw}. 

Furthermore, the Hamiltonian of the Levin-Wen string net model aligns with the constraint equations in gravity. The Wheeler-DeWitt equations adopts the form of difference equations,  fulfilled in our construction thanks to the gluing rules for the quantum $6j$ symbols\cite{Bonzom:2011hm}. The full CFT path integral contains a sum over geometries represented by the integral $\int_0^\infty dP_{\alpha_i}$, and includes all possible geodesic length configurations. Together with the non-perturbative nature of retriangulation invariance, it ensures background independence. We want to emphasize that this \emph{defines} an unconventional non-perturbative Quantum Gravity theory. The BCFT data which is the non-perturbative solution to conformal bootstrap equations specifies the measure for the dual gravitational path integral, dictating the precise weighted sum over quantum geometries necessary for exact holographic duality.

Liouville theory clearly does not produce Quantum Gravity in a conventional manner. It features a flat physical spectrum, in which the vacuum is excluded. This is manifested as a lack of matter sectors in the gravity theory, the latter alone being thermodynamical \cite{Martinec:1998wm, Carlip:1998qw}. However, applying the same construction to a more conventional 2D holographic CFT with a vacuum state and discrete physical Cardy spectrum\cite{Cardy:1986ie, Maloney:2007ud} should yield a dual theory containing a more interesting quantum gravity theory. Unfortunately, non-perturbative formulas for the BCFT structure coefficients of such candidate theories are currently unknown. Nevertheless, it is expected that asymptotic behaviors of $6j$ symbols for holographic CFTs are similar to $\eqref{eq:classlim2}$ for the universal emergence of geometry. 

On the other hand, due to the universality of the high-energy data in chaotic QFTs, which is essentially the eigenstate thermalization hypothesis (ETH) \cite{PhysRevA.43.2046, Srednicki:1994mfb}, we can actually see the universal emergence of geometries from ensemble averages\cite{Saad:2019lba, Belin:2020hea, Chandra:2022bqq, Belin:2023efa}. For instance, with an ETH Ansatz for the averaged BCFT structure coefficients for heavy states above the black hole threshold\cite{Kusuki:2021gpt, Numasawa:2022cni}, thermal AdS and BTZ black holes appear universally in the calculation of torus partition functions\cite{Maloney:2007ud, toappear}. 

Throughout this paper, our primary focus has been on establishing the correspondence between CFT partition functions and gravitational spacetimes. Alternatively, we can explore CFT path integrals with boundaries, which leads to quantum wave-functions\cite{Chen:2024unp}. The holographic tensor network representation can be readily obtained by extending these 2D manifolds into a 3D state-sum. This bridges the Ryu-Takayanagi entanglement entropy formula \cite{Ryu:2006bv} with tensor network as envisaged in \cite{Swingle:2009bg, Miyaji:2016mxg, Caputa:2017urj, Milsted:2018yur, VanRaamsdonk:2018zws}. 

\vspace{-0.8 cm}
\section*{Conclusions}
\vspace{-0.3 cm}
The inspiration for building up spacetimes from basic building blocks went all the way back to Grothendieck, who first coined the term  {\it Lego-Teichm\"uller Game} \cite{lego}, which builds up 2D surfaces from closed pairs of pants. In this essay, we have shown how to build up 2D surfaces and 3D quantum spacetimes using open pairs of pants, which we have named BCFT Legos, as a tribute. Synthesizing ideas from conformal bootstrap, topological symmetries, tensor networks, real-space RG flow, and the asymptotics of quantum $6j$ symbols, we have developed a recipe that churns out precise, non-perturbative and computer friendly 3D path integrals from BCFT data. This framework is broadly applicable, and in the example of Liouville theory, it realises the long-held vision of generating geometry from correlation in an amazingly explicit way. 

Where does that leave us? Constructing non-perturbative examples is undoubtedly challenging, but we hope this example illustrates how the operator algebra of a lower dimensional CFT can universally encode geometries in a higher dimensional theory. The construction seems to be explaining quite generally the origin of the deep connection between gravity/geometry and CFT/algebra. We could for example extend the construction to theories whose structure coefficients exhibit asymptotic behaviors matching spacetimes beyond hyperbolic geometries. The universal emergence of geometry can also be understood in light of the ETH and an ensemble average of the BCFT Legos. 

Could these ideas lead us to more general and interesting quantum gravity theories in 3D and higher dimensions? In 3D, it is perhaps a question of discovering more general module categories. Some work has also been done to generalize this construction to 3D CFT/4D TQFT \cite{Chen:2022wvy}. So far, the 4D bulk theory remains very distinct from gravity, which is not expected to be topological. Nonetheless, a comprehensive description of the bulk necessitates understanding of all the topological symmetries of the 3D CFT, which is itself a very challenging task. One would expect, much like the case of Liouville theory, any complete description of topological symmetries in higher dimensions should also be irrational generally, demanding a deepened understanding of higher fusion tensor categories. It is not inconceivable that some sectors of gravity are hidden in the labyrinth of higher symmetries. Perhaps gravity in general is indeed an exotic phase of quantum matter in a precise sense, as the proverb goes, {\it It From Qubit}. 

\vspace{-0.7 cm}
\section*{Acknowledgments}
\vspace{-0.3 cm}
We thank Ning Bao, Wan Zhen Chua, Zohar Komargodski, Zhihan Liu, Herman Verlinde and Mengyang Zhang for comments and suggestions on the draft. We thank Gong Cheng, Lin Chen, Zheng-Cheng Gu and Bingxin Lao for collaborations on related projects. We thank Zhenhao Zhou for help with plotting the 3D diagrams. LYH acknowledges the support of NSFC (Grant No.11922502, 11875111). YJ acknowledges the support by Novel Quantum Algorithms from Fast Classical Transforms, the U.S Department of Energy ASCR EXPRESS grant, and Northeastern University.
\small
\vspace{-0.3 cm}

\begin{spacing}{1}
\bibliographystyle{ourbst}
\bibliography{gravity}
\end{spacing}
\end{document}